\documentclass{aastex631}

\newcommand{\ha}{\hbox{H$\alpha$}}

\begin{document}

\title{The large-scale structure supplies the formation of gas-star misaligned galaxies}

\author[0009-0005-9342-9125]{Min Bao}
\affiliation{School of Astronomy and Space Science, Nanjing University, Nanjing 210023, China}
\affiliation{Key Laboratory of Modern Astronomy and Astrophysics (Nanjing University), Ministry of Education, Nanjing 210023, China}

\author[0000-0003-3226-031X]{Yanmei Chen}
\affiliation{School of Astronomy and Space Science, Nanjing University, Nanjing 210023, China}
\affiliation{Key Laboratory of Modern Astronomy and Astrophysics (Nanjing University), Ministry of Education, Nanjing 210023, China}

\author[0000-0002-3890-3729]{Qiusheng Gu}
\affiliation{School of Astronomy and Space Science, Nanjing University, Nanjing 210023, China}
\affiliation{Key Laboratory of Modern Astronomy and Astrophysics (Nanjing University), Ministry of Education, Nanjing 210023, China}

\author[0000-0002-4911-6990]{Huiyuan Wang}
\affiliation{Department of Astronomy, University of Science and Technology of China, Hefei 230026, China}
\affiliation{Key Laboratory for Research in Galaxies and Cosmology, University of Science and Technology of China, Hefei 230026, China}

\author[0000-0002-8614-6275]{Yong Shi}
\affiliation{School of Astronomy and Space Science, Nanjing University, Nanjing 210023, China}
\affiliation{Key Laboratory of Modern Astronomy and Astrophysics (Nanjing University), Ministry of Education, Nanjing 210023, China}

\author[0000-0003-2504-3835]{Peng Wang}
\affiliation{Shanghai Astronomical Observatory, Chinese Academy of Sciences, Nandan Road 80, Shanghai 200030, China}

\correspondingauthor{Yanmei Chen}
\email{chenym@nju.edu.cn}

\begin{abstract}

Using the integral field unit data from the Mapping Nearby Galaxies at Apache Point Observatory (MaNGA) survey, we build a sample of gas-star misaligned galaxies. The large-scale environment of misaligned galaxies is dominated by filaments and clusters, while is less dense relative to the gas-star aligned control galaxies. The direction of the large-scale structure (LSS) is defined by its minor axis ($\vec{e_{3}}$), which indicates the slowest collapsing direction. For the aligned controls, the gas and stellar spins are preferentially perpendicular to $\vec{e_{3}}$, since these galaxies reside in high-mass host haloes. For the misaligned galaxies, the gas spins also tend to be perpendicular to $\vec{e_{3}}$, suggesting that misaligned gas is recently accreted from the LSS. Meanwhile, there is no correlation between their stellar spins and $\vec{e_{3}}$. There are two possible explanations for this observational phenomenon: (1) the large-scale environments of misaligned galaxies evolve as they grow, with stellar angular momenta acquiring in different environments having different orientations; (2) the correlation between stellar spins and the LSS is smeared out since a relatively higher portion of misaligned galaxies in sheet environments are statistically analysed together with those in filament environments.

\end{abstract}


\section{Introduction} \label{sec:intro}

Over the past decades, numerical simulations and large redshift surveys have highlighted the large-scale structure (LSS) in the Universe \citep{1996Natur.380..603B, 2015ARA&A..53...51S}. From low to high density, the LSS can be classified into voids, sheets, filaments and clusters. Voids are surrounded by sheets, filaments are at the intersection of sheets, and clusters are at two ends of filaments \citep{2007MNRAS.375..489H}. Large-scale cosmic flows are driven by the buildup of the LSS, with matter migrating from low-density regions to high-density regions \citep{2012MNRAS.427.3320C}. Around each position in the LSS, the direction of cosmic flow can be described by three orthogonal axes. The major axis represents the direction of the fastest collapse, the minor axis represents the direction of the slowest collapse, and the intermediate axis has a collapse speed between them \citep{1970A&A.....5...84Z}. Therefore, the convergence and divergence of cosmic flows along three axes can indicate the density. For a void, the cosmic flows are divergent along all the axes. For a sheet, the cosmic flow along the major axis is convergent, while the flows along the other two axes are divergent. For a filament, the cosmic flows are convergent along the major and intermediate axes, while divergent along the minor axis. For a cluster, the cosmic flows along all the axes are convergent \citep{2012MNRAS.420.1809W}.

In the current cold dark matter cosmogony, the assembly of dark matter haloes occurs with the buildup of the LSS. Numerical simulations have revealed the correlation between halo spins and the minor axis of LSS ($S_h$-LSS correlation), and the dependence of $S_h$-LSS correlation on halo mass (e.g. \citealt{2009ApJ...706..747Z, 2011MNRAS.413.1973W, 2013MNRAS.428.2489L, 2014MNRAS.440L..46A, 2014MNRAS.443.1090F, 2015MNRAS.446.2744L, 2018MNRAS.481..414G}). The spins of low-mass haloes tend to be parallel with the minor axis of LSS, while the spins of high-mass haloes tend to be perpendicular to the minor axis of LSS. Different numerical simulations gave different critical mass (ranging from $M_h \sim 5\times10^{11}h^{-1}~M_\odot$ to $M_h \sim 5\times10^{12}h^{-1}~M_\odot$), at which the $S_h$-LSS correlation changes from parallel to perpendicular orientation. Voids can be regarded as regions devoid of haloes, sheets typically host the low-mass haloes, while the high-mass haloes primarily reside in filaments and clusters \citep{2014MNRAS.441.2923C}. Based on this picture, the transition of parallel to perpendicular $S_h$-LSS correlations can be explained by the following scenarios \citep{2017MNRAS.468L.123W, 2018MNRAS.473.1562W}. On the one hand, the majority of mass assembly of low-mass haloes occurs within sheets, where the matter migrations are perpendicular to the filaments at their intersection (Figure 12 of \citealt{2015MNRAS.452.3369C}). As a result, the low-mass haloes acquire angular momentum parallel with the filaments, which can convert into spins parallel with the minor axis of LSS. On the other hand, mass assembly of high-mass haloes primarily occurs in filaments (including clusters), where the matter migrations are along the filaments (Figure 12 of \citealt{2015MNRAS.452.3369C}). The high-mass haloes obtain angular momentum perpendicular to the filaments, ultimately converting into spins perpendicular to the minor axis of LSS.

Galaxies reside in dark matter haloes, and their formation is connected to the assembly of haloes in which they form \citep{2018ARA&A..56..435W}. Hence, we expect that similar correlations also exist between galaxy spins and the minor axis of LSS ($S_g$-LSS correlation). Using the state-of-the-art hydrodynamic simulation Illustris-1, \cite{2018ApJ...866..138W} investigated the $S_g$-LSS correlation in the local Universe. The spins of low-mass galaxies are preferentially parallel with the minor axis of LSS, while the spins of high-mass galaxies tend to be perpendicular to the minor axis of LSS, in consistent with the behavior of haloes. \cite{2015ApJ...798...17Z} targeted the Sloan Digital Sky Survey (SDSS) galaxies with host halo mass higher than $10^{11.5}h^{-1}~M_\odot$, and found perpendicular $S_g$-LSS correlations. In addition, \cite{2022MNRAS.516.3569B} studied the $S_g$-LSS correlation as functions of various galaxy properties for Sydney-AAO Multi-object Integral Field Spectrograph survey (SAMI), and discovered that bulge mass is the dominated parameter in the transition of parallel to perpendicular $S_g$-LSS correlations. Except bulge mass, other parameters, such as stellar mass, visual morphology, kinematic feature, stellar age, star formation activity and local environment, can also affect the $S_g$-LSS correlation, which were subsequently confirmed by later studies \citep{2021MNRAS.504.4626K, 2023MNRAS.526.1613B, 2015ApJ...798...17Z}. The bulge components can be proper tracers of galaxy mergers, which primarily happen along the filaments and contribute to perpendicular $S_g$-LSS correlation.

Kinematically misaligned galaxies are defined as hosting two components (gas and/or stars), which spin in obviously different directions. The misalignments can exist between gas-gas, gas-star and star-star \citep{2014ASPC..486...51C}. The gas-star misalignment is most universal among the three types, which stands 20-50\% in the quiescent galaxies \citep{2006MNRAS.366.1151S, 2011AAS...21742202D, 2014A&A...568A..70B}, and 2-5\% in the star-forming galaxies \citep{2016NatCo...713269C, 2019MNRAS.483..458B}. This misalignment is believed to form through external processes, dominated by gas accretion \citep{2011MNRAS.413.1373W, 2012A&A...544A..68L, 2015MNRAS.448.1271L, 2022MNRAS.515.5081Z, 2022MNRAS.511.4685X}, which delivers gas with spins misaligned relative to the stellar component. However, it is still an open question whether the misaligned gas is accreted from the LSS or gas-rich satellites. If the accreted gas originates from the LSS, we would expect a correlation between the spins of misaligned gas and the minor axis of LSS. To explore this question, we build a sample of gas-star misaligned galaxies (misaligned galaxies for short) from the MaNGA survey. The MaNGA IFU data and sample selection are introduced in Section \ref{sec:data}. We present the large-scale environments of the misaligned galaxies in Section \ref{sec:LSS}, and compare them with gas-star aligned control galaxies (aligned controls for short). In Section \ref{sec:spin}, we study the $S_g$-LSS correlations for gas and stellar components, and make comparisons between misaligned galaxies and aligned controls.

\section{Data} \label{sec:data}

The data products applied in this study are obtained from the final data release of the MaNGA survey, which includes 10010 unique galaxies. For each MaNGA galaxy, the global properties, including right ascension (RA), declination (Dec), effective radius ($Re$) and axial ratio ($b/a$), are extracted from NASA-Sloan Atlas catalog (NSA, \citealt{2011AJ....142...31B}). The distortion-corrected redshift ($z_{c}$) is obtained from \cite{2014A&A...566A...1T} catalog. The global stellar mass ($M_{\star}$) is extracted from MaNGA data reduction pipeline (DRP, \citealt{2016AJ....152...83L}) catalog. The spatially resolved properties, including gas and stellar velocities, are obtained from MaNGA data analysis pipeline (DAP, \citealt{2019AJ....158..231W}). The sky-subtracted and flux-calibrated 3D spectra are obtained from MaNGA DRP. We stack the spectra within $1.5~Re$, and calculate the global 4000~\AA~break (D$_{n}4000$) as the flux ratio between two narrow bands of 3850-3950~\AA~and 4000-4100~\AA. Moreover, we fit the kinematic position angles of gas ($\phi_{\rm gas}$) and stellar ($\phi_{\star}$) velocity fields using the python-based package \texttt{PAFIT}. The position angle is defined as a counterclockwise angle between north and a line bisecting the velocity field. We only fit the spaxels with {\ha} signal-to-noise ratios (S/N) higher than 3 for gas velocity field, and the spaxels with median spectral S/N higher than 3 for stellar velocity field.

Following the method in \cite{2022MNRAS.515.5081Z}, we build a sample of 496 misaligned galaxies from MaNGA galaxies. We refer readers to \cite{2022MNRAS.515.5081Z} for more details about the sample selection, and only describe the critical criteria here: (1) exclude `lineless' galaxies with {\ha} S/N lower than 3 for more than 90\% spaxels within 1.5~$Re$; (2) select candidates of misaligned galaxies having position angle offset $\Delta\phi\equiv|\phi_{\rm gas} - \phi_{\star}|$ higher than 30$^{\circ}$, with robust position angle measurements (i.e. $\phi_{\rm error} \leq 20^{\circ}$); (3) visually inspect the candidates, removing ongoing mergers and galaxy pairs.

The large-scale environment around a galaxy is expected to shape the intrinsic properties of it \citep{2018MNRAS.474..547K}. In the next section, we will study the $S_g$-LSS correlation for misaligned galaxies, to figure out the origin of misaligned gas (from the LSS or satellites). The LSS characteristics are extracted from the Galaxy Environment for MaNGA Value Added Catalog (GEMA-VAC), which is complete in the redshift range of $z_c \sim [0.01, 0.12]$. We therefore exclude galaxies that are not within this redshift range, leaving 7394 MaNGA galaxies and 395 misaligned galaxies for the following analysis. In order to explore the large-scale environment of misaligned galaxies, we build a control sample of aligned galaxies for comparison. For each misaligned galaxy, we randomly select five aligned controls with similar stellar mass ($|\Delta \log M_{\star}| < 0.1$), stellar population age ($|\Delta$D$_{n}4000| < 0.05$) and redshift ($|\Delta z| < 0.03$) from MaNGA galaxies.

\section{Large-scale environment} \label{sec:LSS}

\subsection{Distance to filament spine} \label{sec:disperse}

\begin{figure*}[h]
     \centering\resizebox{0.95\textwidth}{!}{\includegraphics{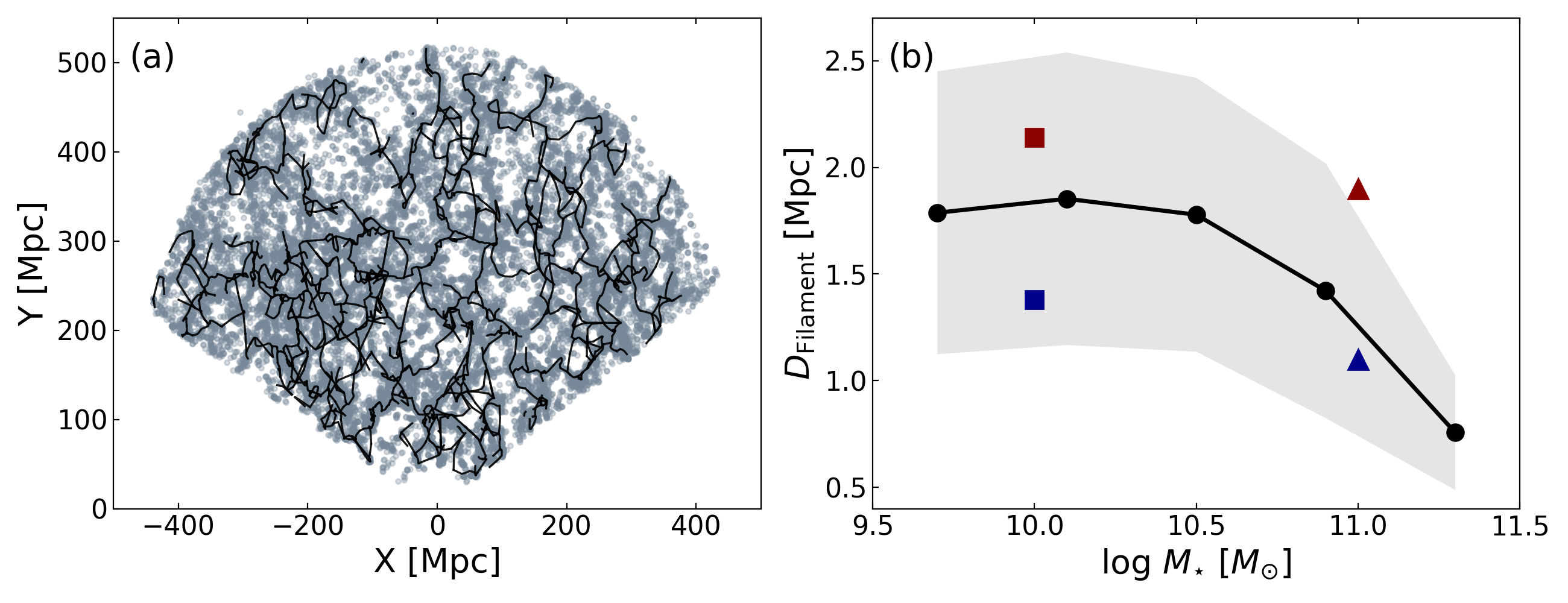}}
    \caption{The distances of galaxies to their nearest filament spines in the filament environments. (a) An illustration of the reconstructed filaments with $\rm Dec \sim[25^{\circ}, 30^{\circ}]$. The grey points show the distributions of SDSS galaxies. The black profiles represent the projected filament spines. (b) The black solid profile shows the $D_{\rm filament}$-$M_{\star}$ correlation for MaNGA galaxies. Each black circle shows the median distance in a certain stellar mass bin. The grey-shaded area shows the 40\%-60\% scattering region. The red square and triangle show the median $D_{\rm filament}$ for low-mass and high-mass misaligned galaxies, while the blue square and triangle show the median $D_{\rm filament}$ for low-mass and high-mass aligned controls, respectively.}
    \label{fig:filament}
\end{figure*}

We reconstruct the 3D filament LSS using a public code named \texttt{DisPerSE} \citep{2011MNRAS.414..350S, 2011MNRAS.414..384S}, which applies a scale-free and parameter-free topological algorithm on the 3D discrete distributions of galaxies. \texttt{DisPerSE} stands for Discrete Persistent Structures Extractor, and is developed for the study of the properties of filament structures in the comic web of galaxy distribution over large scales in the Universe. Taking RA, Dec and $z_{c}$ of SDSS galaxies \citep{2014A&A...566A...1T} as input parameters, we run the \texttt{DisPerSE} code with a 3$\sigma$ persistence threshold, which requires the signals of filament structures at least three times higher than the local noise levels. Figure \ref{fig:filament}(a) displays an illustration of the reconstructed filaments with $\rm Dec \sim[25^{\circ}, 30^{\circ}]$. The grey points show the distributions of SDSS galaxies, and the black profiles represent the projected filament spines.

For each MaNGA galaxy, we define $D_{\rm filament}$ as the smallest 3D Euclidian distance from this galaxy to the points in the nearest filament spine \citep{2022MNRAS.516.3569B}. The radii of filaments have been investigated through both observations and simulations \citep{2005MNRAS.359..272C, 2010MNRAS.409..156B, 2020A&A...638A..75B, 2020A&A...641A.173G, 2021NatAs...5..839W, 2024MNRAS.532.4604W}, which gave typical values spanning from 1~Mpc to 10~Mpc. To explore the correlation between $D_{\rm filament}$ and stellar masses in the filament environments, we only include 4987 galaxies with $D_{\rm filament} \leq 10~\rm Mpc$ from MaNGA galaxies. The black-solid profile in Figure \ref{fig:filament}(b) displays $D_{\rm filament}$-$M_{\star}$ correlation for these galaxies, each black circle shows the median $D_{\rm filament}$ in a certain stellar mass bin. The grey-shaded area shows the 40\%-60\% scattering region. $D_{\rm filament}$ turns out to decrease with increasing stellar mass, which is consistent with previous studies \citep{2018MNRAS.474..547K, 2018MNRAS.474.5437L, 2020MNRAS.491.2864W}. We averagely divide the misaligned galaxies and aligned controls with $D_{\rm filament} \leq 10~\rm Mpc$ into low-mass and high-mass subsamples, based on a cutoff of $\log(M_\ast / M_\odot) = 10.5$. In Figure \ref{fig:filament}(b), the red square and triangle show the median $D_{\rm filament}$ for low-mass and high-mass misaligned galaxies, while the blue square and triangle show the median $D_{\rm filament}$ for low-mass and high-mass aligned controls. MaNGA galaxies, misaligned galaxies and aligned controls exhibit a consistent trend in the filament environments, where $D_{\rm filament}$ decreases as stellar mass increases. This trend can be explained by a scenario that galaxies migrate towards filament spines as they acquire matter through gas accretion or mergers. Along with the migration, increasing density around a galaxy triggers frequent gas accretion or mergers, supplying the mass assembly of this galaxy. Consequently, galaxies with higher stellar mass tend to locate closer to the filament spines.

\subsection{Large-scale environments of misaligned galaxies} \label{sec:environment}

\begin{figure*}[h]
    \centering\resizebox{0.95\textwidth}{!}{\includegraphics{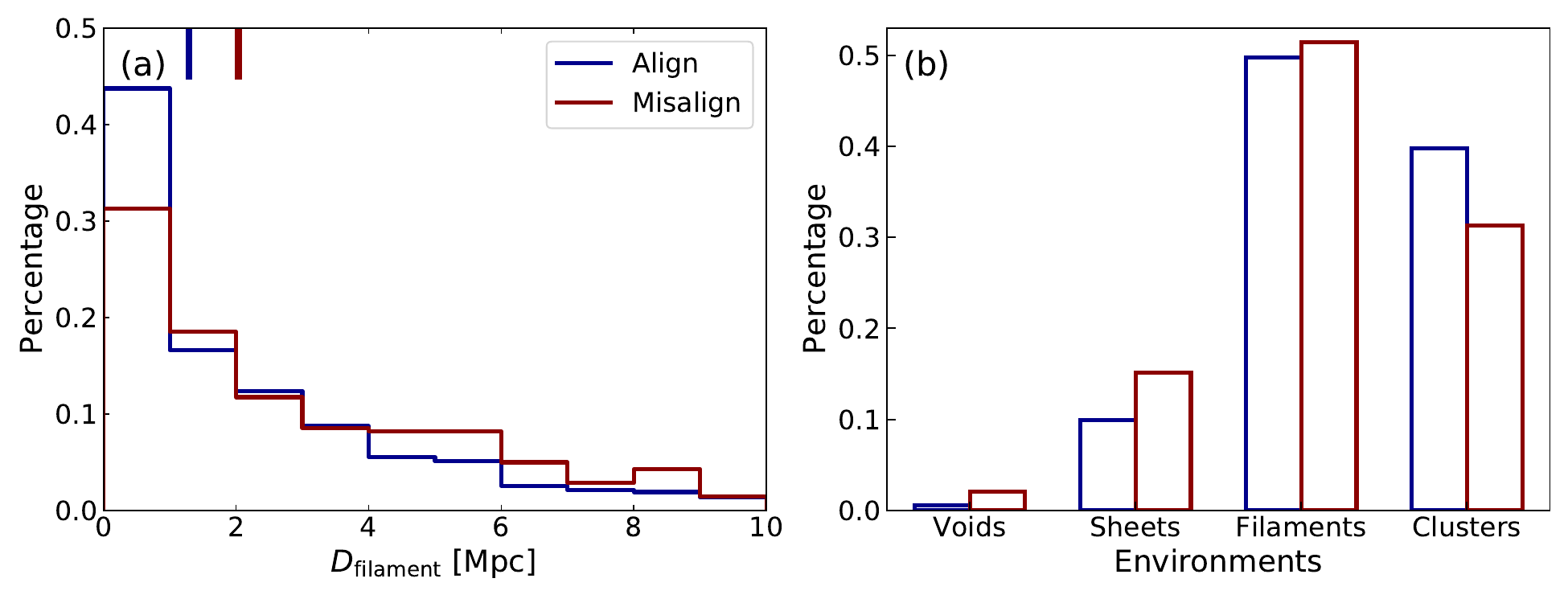}}
   \caption{The large-scale environments of misaligned galaxies and their aligned controls. (a) The red histogram displays the distributions of $D_{\rm filament}$ for misaligned galaxies. The blue histogram displays the distributions of $D_{\rm filament}$ for aligned controls. The red and blue bars on the top show the corresponding median distances. (b) The red histograms display the distributions of environments for misaligned galaxies. The blue histograms display the distributions of environments for aligned controls.}
   \label{fig:environment}
\end{figure*}

Misaligned galaxies are believed to form through external processes, in which their progenitors accrete gas with spins misaligned relative to the stellar component. To figure out the origin of gas accretion (from the LSS or satellites), we make comparisons in large-scale environments between misaligned galaxies and aligned controls. Figure \ref{fig:environment}(a) displays the normalized distributions of $D_{\rm filament}$ for misaligned galaxies (red) and aligned controls (blue) in the filament environments ($D_{\rm filament} \leq 10~\rm Mpc$), respectively. The red and blue bars on the top show the corresponding median values of the distances. The misaligned galaxies locate $\sim$0.75~Mpc farther away from filament spines than their aligned controls, which impies that these galaxies reside in less dense environments.

To quantify the large-scale environments of misaligned galaxies and aligned controls, we collect the LSS characteristics of these galaxies from the GEMA-VAC catalog. In this catalog, the gravitational potential field is provided by ELUCID project (Exploring the Local Universe with the reConstructed Initial Density field, \citealt{2016ApJ...831..164W}), which reconstructed the mass density field based on the galaxy groups in the SDSS survey \citep{2005MNRAS.356.1293Y, 2007ApJ...671..153Y}. The tidal tensor is the Hessian of the gravitational potential. By diagonalizing the tidal tensor at the position of each galaxy, three eigenvalues $T_{1}$, $T_{2}$, $T_{3}$ ($T_{1} > T_{2} > T_{3}$) can be obtained for this galaxy. The eigenvectors $\vec{e_{1}}$, $\vec{e_{2}}$ and $\vec{e_{3}}$ are computed from corresponding eigenvalues, representing the major, intermediate and minor axes of the LSS. The positive and negative eigenvalues indicate the convergence and divergence of cosmic flows along corresponding axes. Different environments can be classified by three eigenvalues: $T_{1}$, $T_{2}$, $T_{3} < 0$ represents voids; $T_{1} > 0$ \& $T_{2}$, $T_{3} < 0$ represent sheets; $T_{1}$, $T_{2} > 0$ \& $T_{3} < 0$ represent filaments; $T_{1}$, $T_{2}$, $T_{3} > 0$ represents clusters.

In Figure \ref{fig:environment}(b), we compare the large-scale environments between misaligned galaxies (red) and aligned controls (blue). On the one hand, the large-scale environments of both misaligned galaxies and aligned controls are dominated by filaments and clusters. On the other hand, more misaligned galaxies reside in voids, sheets and filaments, while less misaligned galaxies reside in clusters compared with their aligned controls. This result is consistent with what we find in Figure \ref{fig:environment}(a), the misaligned galaxies statistically reside in less dense environments. Such a phenomenon was also reported in the previous studies, e.g. in the ATLAS$^{\rm 3D}$ survey \citep{2011MNRAS.417..882D} and in the MaNGA survey \citep{2016MNRAS.463..913J}.

\section{Formation of misaligned galaxies} \label{sec:spin}

Taking advantage of the spatially resolved velocity fields, we compute the spin vectors of gas and stellar components for each MaNGA galaxy following the method in \cite{2015ApJ...798...17Z}. Firstly, the inclination angle $\zeta$ between the plane of galactic disk and the line of sight is calculted through
\begin{equation}
    \sin^{2}\zeta = \frac{(b/a)^{2} - f^{2}}{1 - f^{2}},
    \label{e1}
\end{equation}
where the intrinsic axial ratio ($f$) is set to 0.14 following \cite{2012ApJ...744...82V}. We set $\zeta = 0$ for galaxies with $b/a < f$. Secondly, the spin vector is given by 
\begin{eqnarray}
    S_{x} = \cos \alpha \cos \delta \sin \zeta + \cos \zeta (\sin \phi \cos \alpha \sin \delta - \cos \phi \sin \alpha), \label{e2} \\
    S_{y} = \sin \alpha \cos \delta \sin \zeta + \cos \zeta (\sin \phi \sin \alpha \sin \delta + \cos \phi \cos \alpha), \label{e3} \\
    S_{z} = \sin \delta \sin \zeta - \cos \zeta \sin \phi \cos \delta, \label{e4}
\end{eqnarray}
where $\alpha$ and $\delta$ are RA and Dec, $\zeta$ is calculated through Equation (\ref{e1}), and $\phi$ is taken as $\phi_{\rm gas}$ and $\phi_{\star}$ for spin vectors of gas and stellar components, respectively.

To describe the $S_g$-LSS correlation, we calculate the absolute value of the cosine angle between spin vector and $\vec{e_{3}}$ of the LSS at the position of each galaxy, which is given by
\begin{equation}
    |\cos \theta| = \frac{|\vec{S} \cdot \vec{e_{3}}|}{|\vec{S}| \cdot |\vec{e_{3}}|},
    \label{e5}
\end{equation}
where $\vec{S}$ is the spin vector of gas or stellar component. The gas (or stellar) spin is parallel with $\vec{e_{3}}$ when $|\cos \theta| \sim 1$, while the gas (or stellar) spin is perpendicular to $\vec{e_{3}}$ when $|\cos \theta| \sim 0$.

Following \cite{2015ApJ...798...17Z}, we generate 100 mock samples, each containing the same number of galaxies as the MaNGA sample, to quantify the significance of $S_g$-LSS correlation for MaNGA galaxies. In each mock sample, $\vec{e_{3}}$ directions are kept the same as the relevant MaNGA galaxies, but the spin vectors of the mock galaxies are randomly distributed. The normalized probability distribution function (PDF) of $|\cos \theta|$ is calculated through
\begin{equation}
    P(|\cos \theta|) = \frac{N(|\cos \theta|)}{\langle N_{m}(|\cos \theta|) \rangle},
    \label{e6}
\end{equation}
where $N(|\cos \theta|)$ is the number of MaNGA galaxies in each $|\cos \theta|$ bin, and $\langle N_{m}(|\cos \theta|) \rangle$ is the average number of mock galaxies in each $|\cos \theta|$ bin among 100 mock samples. We also calculate the normalized standard deviation ($\pm 1\sigma$ scatter) from 100 mock samples as $S(|\cos \theta|) = \sigma_m(|\cos \theta|)/\langle N_{m}(|\cos \theta|) \rangle$ to assess the significance of correlation, where $\sigma_m(|\cos \theta|)$ is the average standard deviation of $N_{m}(|\cos \theta|)$ obtained from 100 mock samples. Using a similar method, we calculate the $\pm 2\sigma$ and $\pm 3\sigma$ scatters from 100 mock samples.

\begin{figure*}[h]
    \centering\resizebox{1\textwidth}{!}{\includegraphics{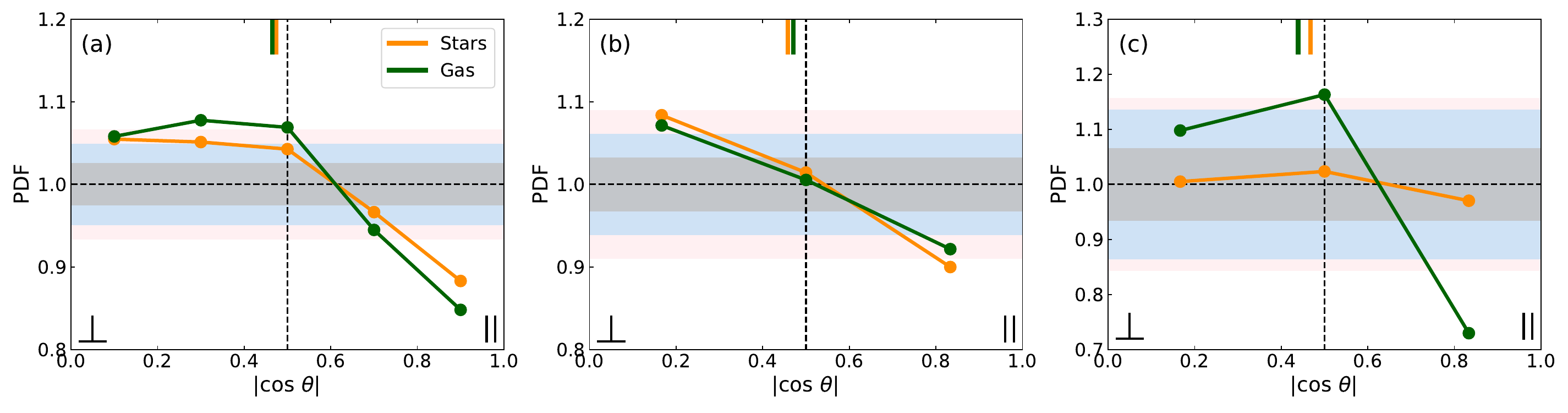}}
   \caption{The distributions of $|\cos \theta|$, where $\theta$ is the angle between gas (or stellar) spins and $\vec{e_{3}}$ of the LSS. (a) The PDFs of $|\cos \theta|$ for MaNGA galaxies. (b) The PDFs of $|\cos \theta|$ for aligned controls. (c) The PDFs of $|\cos \theta|$ for misaligned galaxies. In panels (a), (b) \& (c), the green circles show the PDFs of $|\cos \theta|$ between gas spins and $\vec{e_{3}}$, the orange circles show the PDFs of $|\cos \theta|$ between stellar spins and $\vec{e_{3}}$. The green and orange bars on the top mark the corresponding median values. The grey-, blue- and pink-shaded areas show the $\pm 1\sigma$, $\pm 2\sigma$ and $\pm 3\sigma$ scatters for the corresponding mock samples.}
   \label{fig:spin_manga}
\end{figure*}

Figure \ref{fig:spin_manga}(a) displays the PDFs of $|\cos \theta|$ for MaNGA galaxies, with green and orange lines representing gas and stellar components, respectively. The green and orange bars on the top of Figure \ref{fig:spin_manga}(a) mark the median values of $|\cos \theta|$ for gas and stellar components. It turns out that both the gas and stellar spins of MaNGA galaxies tend to be perpendicular to $\vec{e_{3}}$ with $P(|\cos \theta| < 0.5) > 1$. The grey-, blue- and pink-shaded areas in Figure \ref{fig:spin_manga}(a) show the $\pm 1\sigma$, $\pm 2\sigma$ and $\pm 3\sigma$ scatters from mock samples. It is obvious that the perpendicular signals for both gas and stellar components approach $3\sigma$ significance in Figures \ref{fig:spin_manga}(a). The perpendicular signals arise because these MaNGA galaxies reside in high-mass host haloes. We extract host halo masses for MaNGA galaxies from the \cite{2007ApJ...671..153Y} catalog, and found all higher than $M_h \sim 5 \times 10^{11}~M_\odot$. The high-mass haloes primarily acquire mass in filaments (including clusters), hence obtain spins perpendicular to the minor axis of the LSS.

Comparing the $S_g$-LSS correlations between misaligned galaxies and aligned controls can help us in understanding the origin of the accreted gas in the misaligned galaxies. Figure \ref{fig:spin_manga}(b) displays the normalized PDFs of $|\cos \theta|$ for aligned controls, with green and orange colors representing gas and stellar components. The green and orange bars on the top show the corresponding median values. Both gas and stellar components in the aligned controls show perpendicular signals, approaching $3\sigma$ significance (pink-shaded area). Figure \ref{fig:spin_manga}(c) displays the normalized PDFs of $|\cos \theta|$ for misaligned galaxies, the lines and symbols are color-coded in the same way as that in Figure \ref{fig:spin_manga}(b). The correlation between gas spins and $\vec{e_{3}}$ presents perpendicular signals that approach $2\sigma$ significance (blue-shaded area). However, there is no correlation between stellar spins and $\vec{e_{3}}$, as the orange line locates within the grey-shaded area.

Numerical simulations and observational studies (e.g. \citealt{2018ApJ...866..138W, 2020MNRAS.491.2864W, 2022MNRAS.516.3569B}) have found evidence of different formation scenarios for low-mass and high-mass galaxies. The majority mass assembly of low-mass galaxies occurs within sheets, hence the galaxy spins are preferentially parallel with the minor axis of LSS. Meanwhile, the mass assembly of high-mass galaxies primarily occurs along filaments, the galaxy spins therefore tend to be perpendicular to the minor axis of LSS. It is within our expectation that the $S_g$-LSS correlations of gas and stellar components in aligned galaxies (Figure \ref{fig:spin_manga}b) are consistent with that in MaNGA galaxies (Figure \ref{fig:spin_manga}a), since the aligned controls are a subset of MaNGA galaxies and reside in high-mass host haloes with $M_h \gtrsim 5 \times 10^{11}~M_\odot$. The perpendicular $S_g$-LSS correlation of gas component in misaligned galaxies agrees with that in aligned controls (Figure \ref{fig:spin_manga}c), suggesting that misaligned gas is recently accreted from the LSS. It is interesting to find that the stellar spins in misaligned galaxies shows no correlation with the LSS. One possibility is that the large-scale environments of misaligned galaxies evolve as they grow, with stellar angular momenta acquiring in different environments having different orientations. Another possibility is that a higher portion of misaligned galaxies, compared with the aligned controls, reside in sheet environments (Figure \ref{fig:environment}b). Since these galaxies are statistically analysed together with misaligned galaxies in filament environments, the $S_g$-LSS correlation of stellar component may be smeared out. To verify these possibilities, we expect to collect data from various IFU surveys, such as SAMI \citep{2012MNRAS.421..872C} and Calar Alto Legacy Integral Field Area (CALIFA) \citep{2012A&A...538A...8S}, to enlarge the sample of misaligned galaxies in the future.

\section{Conclusion} \label{sec:conclusion}
In this study, we build a sample of gas-star misaligned galaxies from the MaNGA survey, as well as an aligned control sample which is closely matched in stellar mass, stellar population age and redshift. We explore the large-scale environments and the correlations between spins and the large-scale structure for misaligned galaxies and aligned controls to understand the origin of accreted gas in the misaligned galaxies. The main results are as follows:

\begin{itemize}

    \item[1.] We study the correlation between distances to the nearest filament spines and stellar masses for MaNGA galaxies in the filament environments. The distances turn out to decrease with increasing stellar masses, which favors a scenario of galaxy mass assembly primarily happening with migration towards filament spines. Compared with the aligned controls, the misaligned galaxies locate farther away from filament spines, implying that they reside in less dense environments.

    \item[2.] From low to high density, the large-scale structure can be classified into voids, sheets, filaments and clusters. The large-scale environments of both misaligned galaxies and aligned controls are dominated by filaments and clusters. Compared with the aligned controls, more misaligned galaxies reside in voids, sheets and filaments, while less misaligned galaxies reside in clusters, which supports the result that misaligned galaxies reside in less dense environments.

    \item[3.] For the aligned controls, the gas and stellar spins are preferentially perpendicular to the slowest collapsing direction ($\vec{e_{3}}$) of the large-scale structure, since these galaxies reside in high-mass host haloes. For the misaligned galaxies, the gas spins also tend to be perpendicular to $\vec{e_{3}}$, suggesting that misaligned gas is recently accreted from the LSS. Meanwhile, there is no correlation between their stellar spins and $\vec{e_{3}}$. One possibility is that the large-scale environments of misaligned galaxies evolve as they grow, with stellar angular momenta acquiring in different environments having different orientations. Another possibility is that the correlation between stellar spins and the LSS is smeared out since a relatively higher portion of misaligned galaxies in sheet environments are statistically analysed together with those in filament environments.

\end{itemize}
     
Acknowledgements: MB acknowledges support by the National Natural Science Foundation of China, NSFC Grant No. 12303009. YMC acknowledges support by the National Natural Science Foundation of China, NSFC Grant Nos. 12333002, 11573013, 11733002, 11922302 and the China Manned Space Project, NO. CMS-CSST-2021-A05. HYW acknowledges support by the National Natural Science Foundation of China, Nos. 12192224 and CAS Project for Young Scientists in Basic Research, Grant No. YSBR-062. PW acknowledges support by the NSFC, No. 12473009 and sponsored by Shanghai Rising-Star Program, No.24QA2711100.

Funding for the Sloan Digital Sky Survey IV has been provided by the Alfred P. Sloan Foundation, the U.S. Department of Energy Office of Science, and the Participating Institutions. SDSS- IV acknowledges support and resources from the Center for High-Performance Computing at the University of Utah. The SDSS web site is www.sdss.org. SDSS-IV is managed by the Astrophysical Research Consortium for the Participating Institutions of the SDSS Collaboration including the Brazilian Participation Group, the Carnegie Institution for Science, Carnegie Mellon University, the Chilean Participation Group, the French Participation Group, Harvard-Smithsonian Center for Astrophysics, Instituto de Astrof\'{i}sica de Canarias, The Johns Hopkins University, Kavli Institute for the Physics and Mathematics of the Universe (IPMU) / University of Tokyo, Lawrence Berkeley National Laboratory, Leibniz Institut  f\"{u}r Astrophysik Potsdam (AIP), Max-Planck-Institut  f\"{u}r   Astronomie  (MPIA  Heidelberg), Max-Planck-Institut   f\"{u}r   Astrophysik  (MPA   Garching), Max-Planck-Institut f\"{u}r Extraterrestrische Physik (MPE), National Astronomical Observatory of China, New Mexico State University, New York University, University of Notre Dame, Observat\'{o}rio Nacional / MCTI, The Ohio State University, Pennsylvania State University, Shanghai Astronomical Observatory, United Kingdom Participation Group, Universidad Nacional  Aut\'{o}noma de M\'{e}xico,  University of Arizona, University of Colorado  Boulder, University of Oxford, University of Portsmouth, University of Utah, University of Virginia, University  of Washington,  University of  Wisconsin, Vanderbilt University, and Yale University.

\end{document}